\documentclass[aps,prl,showpacs,noshowkeys,amsmath,amssymb,amsfonts,reprint]{revtex4-1}
\usepackage{graphicx}
\usepackage{dcolumn}
\usepackage{bm}
\usepackage{color}
\usepackage{soul}
\usepackage[normalem]{ulem}

\begin{document}

\title{Self-thermophoresis at the nanoscale using light induced solvation dynamics}

\author{Carles Calero$^{1,2}$}
\author{Edwin L. Sibert III$^3$}
\author{Rossend Rey$^4$}
\affiliation{
$^1$Departament de F\'{i}sica de la Mat\`{e}ria Condensada, Universitat de Barcelona, Barcelona, Spain\\
$^2$Institut de Nanoci\`{e}ncia i Nanotecnologia, Universitat de Barcelona, Barcelona, Spain\\
$^3$Department of Chemistry and Theoretical Chemistry Institute, University of Wisconsin-Madison, Madison, Wisconsin 53706, USA\\
$^4$Departament de F\'{\i}sica, Universitat Polit\`{e}cnica de Catalunya, Campus Nord B4-B5, Barcelona 08034, Spain.}

\date{\today}

\begin{abstract}
Downsizing microswimmers to the nanoscale, and using light as an externally controlled fuel, are two important goals within the field of active matter. 
Here we demonstrate using all-atom molecular dynamics simulations that solvation relaxation, the solvent dynamics induced after visible light electronic excitation of a fluorophore, can be used to propel nanoparticles immersed in polar solvents. We show that fullerenes functionalized with fluorophore molecules in liquid water exhibit substantial enhanced mobility under external excitation, with a propulsion speed proportional to the power dissipated into the system. We show that the propulsion mechanism is quantitatively consistent with a molecular scale instance of self-thermophoresis. Strategies to direct the motion of functionalized fullerenes in a given direction using confined environments are also discussed. 
\end{abstract}

\maketitle
Inspired by the study of self-propulsion in microbiology~\cite{purcell:1977}, there is a strong interest in the design of artificial active particles (``swimmers'')~\cite{kapral:2013,elgeti:2015,bechinger:2016,moran:2017,xu:2017,xu:2017bis}. The current flurry of propulsion mechanisms~\cite{bechinger:2016} arises from both fundamental and applied motivations. Achieving control of micro/nano-particles in the liquid phase holds the promise to be truly disruptive in a variety of fields. Although an ample variety of swimmers have been successfully devised and implemented~\cite{bechinger:2016,calero:2018}, substantial hurdles remain. It has been emphasized that downsizing to the nanometer scale is probably the most pressing goal~\cite{bechinger:2016}. A second issue concerns the propulsion mechanism. The most promising mechanisms seem tied to the ability to sustain chemical reactions at the swimmer's surface, a problematic feature for biological applications~\cite{xu:2017,xu:2017bis}. This has prompted the call for swimmers that could rely on electromagnetic radiation~\cite{xu:2017,xu:2017bis}. On top of being more biologically friendly, light activation enables the ability to externally turn propulsion on and off.

Light activated self-thermophoresis was recently observed experimentally in a landmark study of Janus (metal/dielectric) particles \cite{jiang:2010}: a laser beam heats the metallic hemisphere, resulting in an inhomogeneous temperature distribution of the immediate solvent which induces motion of the particle. This effect has also been found for a computational mesoscopic model~\cite{yang:2011}. The question arises of whether a similar feat may be achieved at the molecular scale. Indeed, there is no shortage of mechanisms, so far untapped, by which energy can be absorbed by a molecule and released into the immediate environment~\cite{yardley:1980}. Activation of rotational, vibrational and electronic degrees of freedom, and ascertaining the pathways by which excess energy is channeled are fundamental topics with ongoing interest in the physical chemistry community~\cite{oxtoby:1981,bagchi:1983,rey:1996,ohta1999,sibert:2002,rey:2004,rey:2012}. 

In this Letter we investigate the use of light induced solvation dynamics to propel nanoparticles immersed in polar solvents. We refer to the solvent relaxation that follows (visible light) electronic excitation of fluorophore molecules, which is usually accompanied by a charge redistribution within the solute~\cite{maroncelli:1993,bagchi:2010}. For polar solvents (e.g. water), the perturbed charge distribution induces strong Coulomb forces, producing fast reorientational dynamics of solvent molecules in the solute's vicinity. Germane to this rearrangement there is a concomitant energy flux, from the newly created excess solute-solvent electrostatic energy into solvent kinetic energy. When the electronic state subsequently relaxes, with the redshift of the emitted photon accounting for the energy transferred (Stokes shift), additional solvent rearrangement and energy transfer follows. For the scenario that has been the workhorse in the study of solvation dynamics (atomic ionization in water), energies of $\sim 100$ kcal/mol are transferred on a 100 fs timescale to the librations of the immediate waters~\cite{rey:2015}. Such efficient energy transfer results in strong heating of the fluorophore's surroundings which can be sustained in time by excitation/deexcitation cycles. The power transfer into the solvent is many orders of magnitude higher than previously considered cases (e.g. exothermic calatytic reactions in enzymes~\cite{riedel:2014,wand:2014,golestanian:2015}), and could be used to propel particles at the nanometer scale by self-thermophoresis. 

\begin{figure}[ht]
\includegraphics[height=4cm]{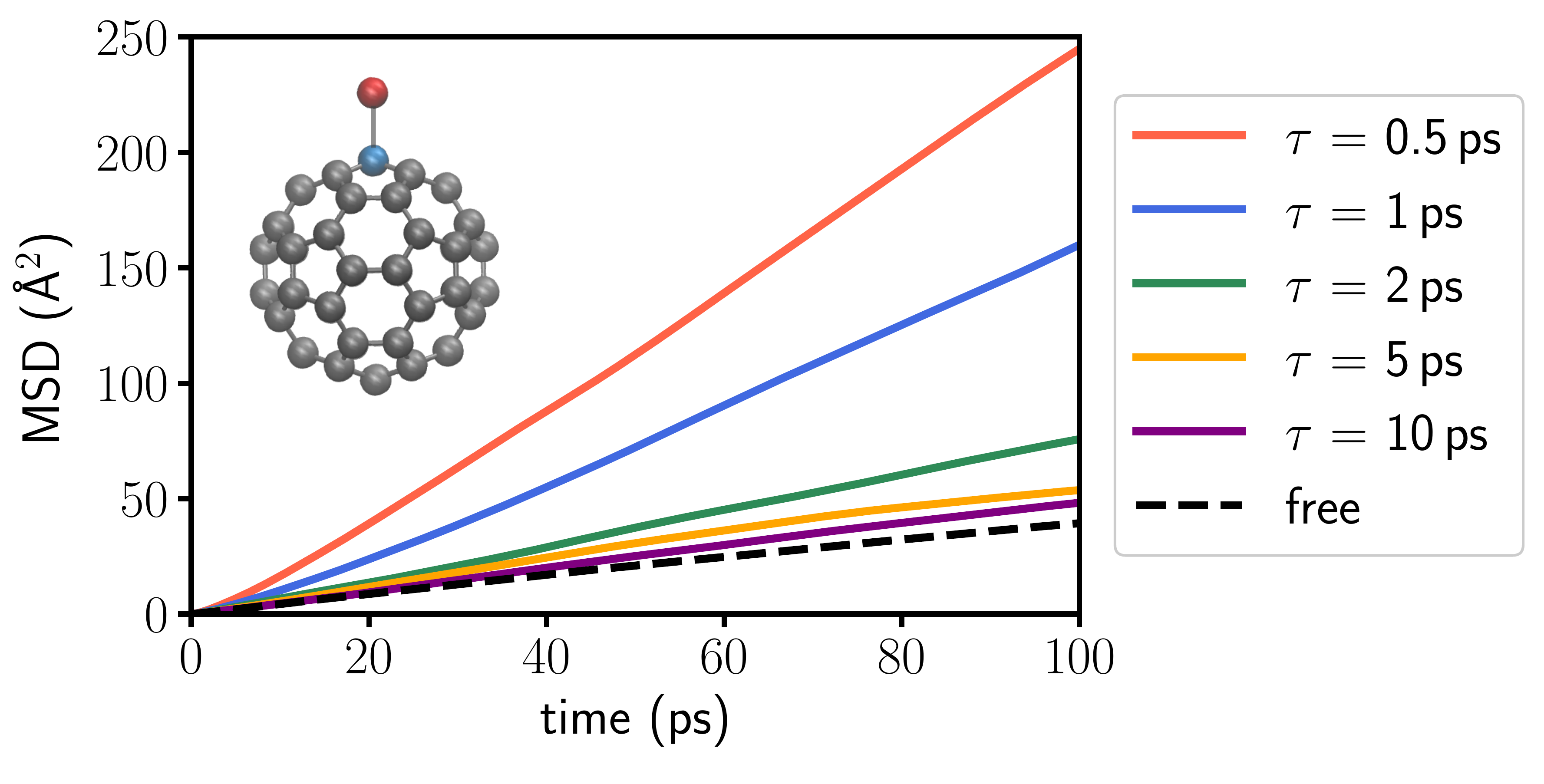}
\caption{Mean square displacement (MSD) of the C$_{60}$-X center of mass for free diffusion (dashed lines) and for different dipole switching periods $\tau = 0.5, 1, 2, 5, 10$ ps. }
\label{fig:diffusive}
\end{figure}

To test these ideas we consider a nanometric propeller composed of a Buckminster fullerene (C$_{60}$) functionalized with a fluorophore and immersed in liquid water. The choice of C$_{60}$ is motivated by its size ($\sim 1$nm) and high degree of symmetry, which should facilitate the assessment of the factors contributing to its mobility and permit comparison with analytic results for spherical bodies. In addition, highly accurate atomic models for C$_{60}$ (and water) are available. Fullerenes can be functionalized with exohedrally attached groups which not only increase their low solubility, but can also change their biological activity~\cite{goodarzil:2017} or, as shown here, can fundamentally alter their mobility. Inspired by metallo-fullerenes~\cite{gonzalez:2017} we have chosen the simplest possibility, namely grafting a single atom X to the C$_{60}$ (see inset in Fig.~\ref{fig:diffusive} where the propeller, called C$_{60}$-X, is schematically shown). The atom attached creates a localized (diatomic) electric dipole C-X~\cite{gonzalez:2017} which can be externally excited~\cite{SM}. Due to the separation in timescales between charge redistribution within the fluorophore after excitation, and the subsequent relaxation of the surrounding solvent, the effect of the external excitation is prescribed as an instantaneous switch of the electric dipole direction, a standard procedure~\cite{bruehl:1992, schile:2017}. The dynamics of the actuated C$_{60}$-X molecule immersed in water is then solved through all-atom molecular dynamics (MD) simulations~\cite{SM}. The effect of the actuation is studied as a function of dipole strength and switching time. 

First, we analyze the free dynamics of the C$_{60}$-X model. Motion occurs at very low Reynolds number assuming typical speeds~\cite{elgeti:2015} ($Re \approx 10^{-6}$), a regime where a spherical particle is characterized by the translational Stokes-Einstein diffusion coefficient $D_t = k_B T/6 \pi \eta R$, and the rotational diffusion coefficient $D_r = k_B T/8 \pi \eta R^3$. 
Although the continuum assumption is not exact at the nanoscale~\cite{hynes:1979}, these formulas are accurate enough provided one allows some leeway for the hydrodynamic radius $R$ (with usually slightly different translational and rotational values). Fig.~\ref{fig:diffusive} displays MD results for the translational mean square displacement (MSD) at a temperature of 300 K, with a diffusion coefficient of 0.75 nm$^2$/ns. Retrieving the water model's viscosity ($\eta= 0.43 $ mPa$\cdot$s for TIP3P-Ew \cite{lai:2010}), the Stokes-Einstein formula produces a radius of $R \approx 6.8$ \AA, while from the rotational MSD we obtain $R \approx 5.3$ \AA. These values are satisfactorily consistent with the fact that water oxygen density peaks at a distance of $\approx 6.6$ {\AA} from C$_{60}$-X molecule center~\cite{SM}. 

We now check whether our model conforms to what is usually described as a hot Brownian particle. The procedure to simulate solvation relaxation classically (see Ref.~\citenum{rey:2016} for a fuller account of the theoretical framework) does not require to directly include the radiation or compute non-adiabatic couplings: at a given time the dipole is simply inverted, causing the response and consequent heating of the immediate environment, followed by a subsequent inversion (relaxation) after an adequately chosen period of time. Here the excitation-relaxation sequence is repeated periodically, with a time $\tau$ between inversions. Fig.~\ref{fig:diffusive} compares the free diffusion MSD to that resulting from several rates of dipole inversion. A core result stands out: under the effect of excitation-relaxation cycles our solute displays substantially enhanced mobility, which increases with power absorbed. 

\begin{figure}[ht]
\includegraphics[height=5cm]{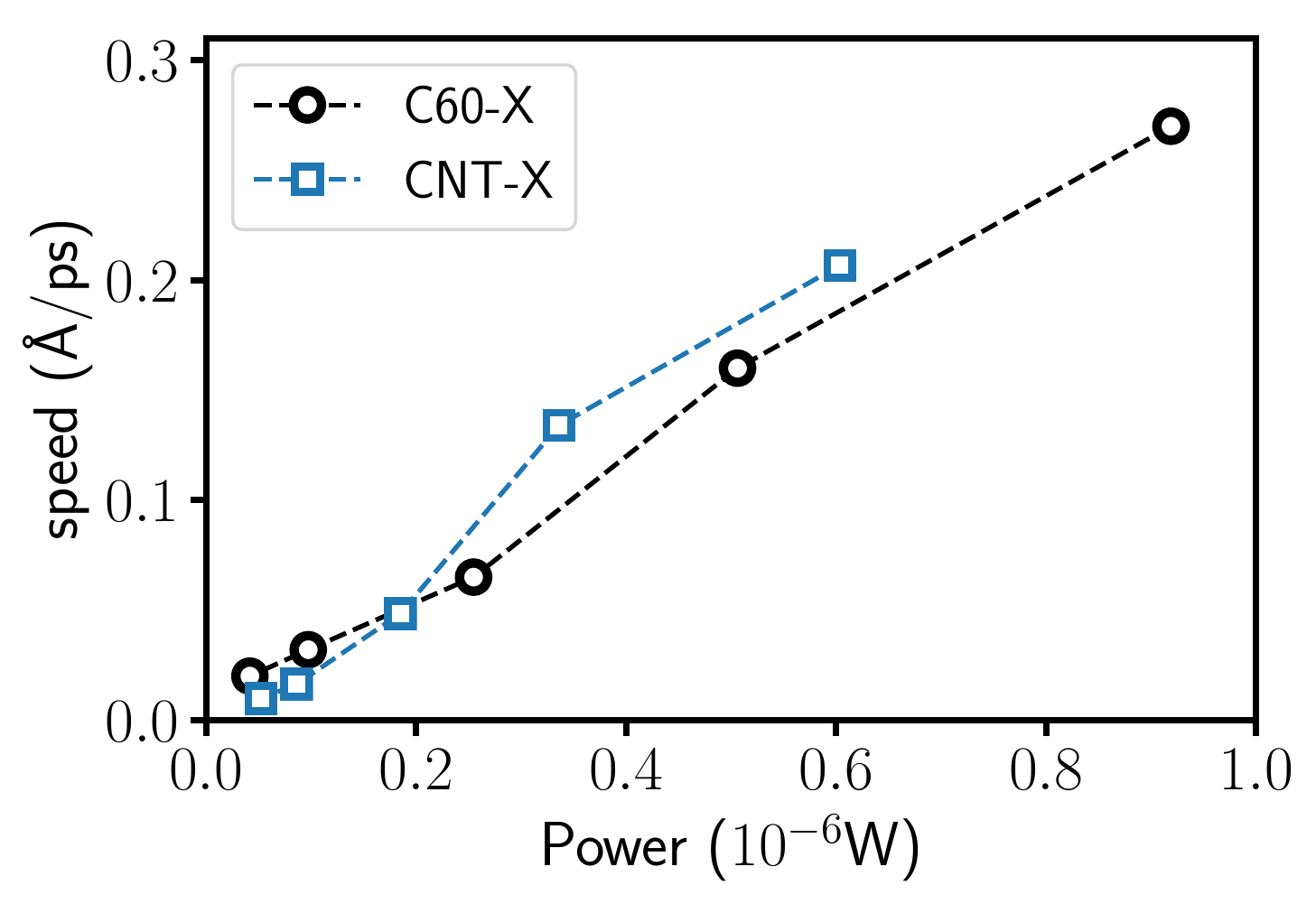}
\caption{Speed of C$_{60}$-X and CNT-X molecules as a function of the power absorbed. The values for C$_{60}$-X result from a multiple regression of the MSD dependence using Eq.~\ref{eq:diffusive}~\cite{SM}. In contrast, the speed of CNT-X is calculated from the net displacement of the molecule before it reaches the edge of the confining nanotube.}
\label{fig:speed}
\end{figure}

For a quantitative estimate, we resort to the standard model~\cite{bechinger:2016} of a self-propelled Brownian particle with propulsion speed $v$. In this model, since the direction of motion is subjected to rotational diffusion, a coupling rotation-translation arises \cite{bechinger:2016}. The following analytic MSD is obtained for a spherical particle in three dimensions (Ref.~\citenum{hagen:2011}, model (3,2,s))

\begin{equation}
\label{eq:diffusive}
\left\langle \Delta \vec{r}^2 \right\rangle = \left[ 6 D_t + \frac{v^2}{D_r} \right]t + \frac{1}{2} \left( \frac{v}{D_r}\right)^2\left[e^{-2 D_r t} - 1\right] .
\end{equation}
We have adjusted Eq.~\ref{eq:diffusive} to the curves in Fig.~\ref{fig:diffusive}~\cite{SM} to obtain a central quantitative result, displayed in  Fig.~\ref{fig:speed}: a linear increase of the propulsion speed with increasing applied power (computed from the energy deposited during the simulation~\cite{SM}). 
This result strongly suggests the presence of a new type of propelling motor, one for which the basic mechanism is the excitation of water librations by means of non-equilibrium Coulomb forces produced by external light excitation, i.e. what could be termed a light-fuelled Coulomb motor. 
 
The linear dependence of the C$_{60}$-X speed on applied power suggests that propulsion can be understood, from a continuum mechanics perspective, as a molecular scale instance of self-thermophoresis~\cite{golestanian:2007,golestanian:2015}. Indeed, the basic (order of magnitude) formulas for thermophoresis of a spherical solute~\cite{golestanian:2015} show that speed depends linearly on the power applied ($P$) and can be estimated using~\cite{golestanian:2007} 
\begin{equation}
\label{eq:speed}
v \sim (D_T /\kappa R^2) P \equiv \alpha \cdot P\,,
\end{equation}
where $\kappa$ is the solvent thermal conductivity ($\kappa \sim 0.8$W/m$\cdot$K for TIP3P-Ew water\cite{sirk:2013}), and $D_T$ stands for the thermal diffusion coefficient of the solute (defined by the limit velocity under a temperature gradient \cite{piazza:2008}, $ \vec{v} = - D_T \nabla T$). The precise value of $D_T$ depends on subtle aspects, and a general theory is currently lacking~\cite{piazza:2008}. Remarkably, experimental measures show that $D_T$ does not depend on the particle size and its order of magnitude is rather universal, spanning a very limited range~\cite{piazza:2008}: $10^{-12}<D_T< 10^{-11}$ m$^2$/s$\cdot$K. Using Eq.~\ref{eq:speed} we can thus provide a theoretical prediction for the range spanned by $\alpha$ of $\sim [3\cdot 10^6,3\cdot10^7]$ N$^{-1}$. In comparison, a linear fit of the computational results in Fig.~\ref{fig:speed} yields $\alpha_{sim} \approx 3\cdot 10^7$  N$^{-1}$, which lies within the estimated theoretical range. This notable match strongly supports the self-thermophoretic origin of the propulsion mechanism.

\begin{figure}[ht]
\includegraphics[height=5cm]{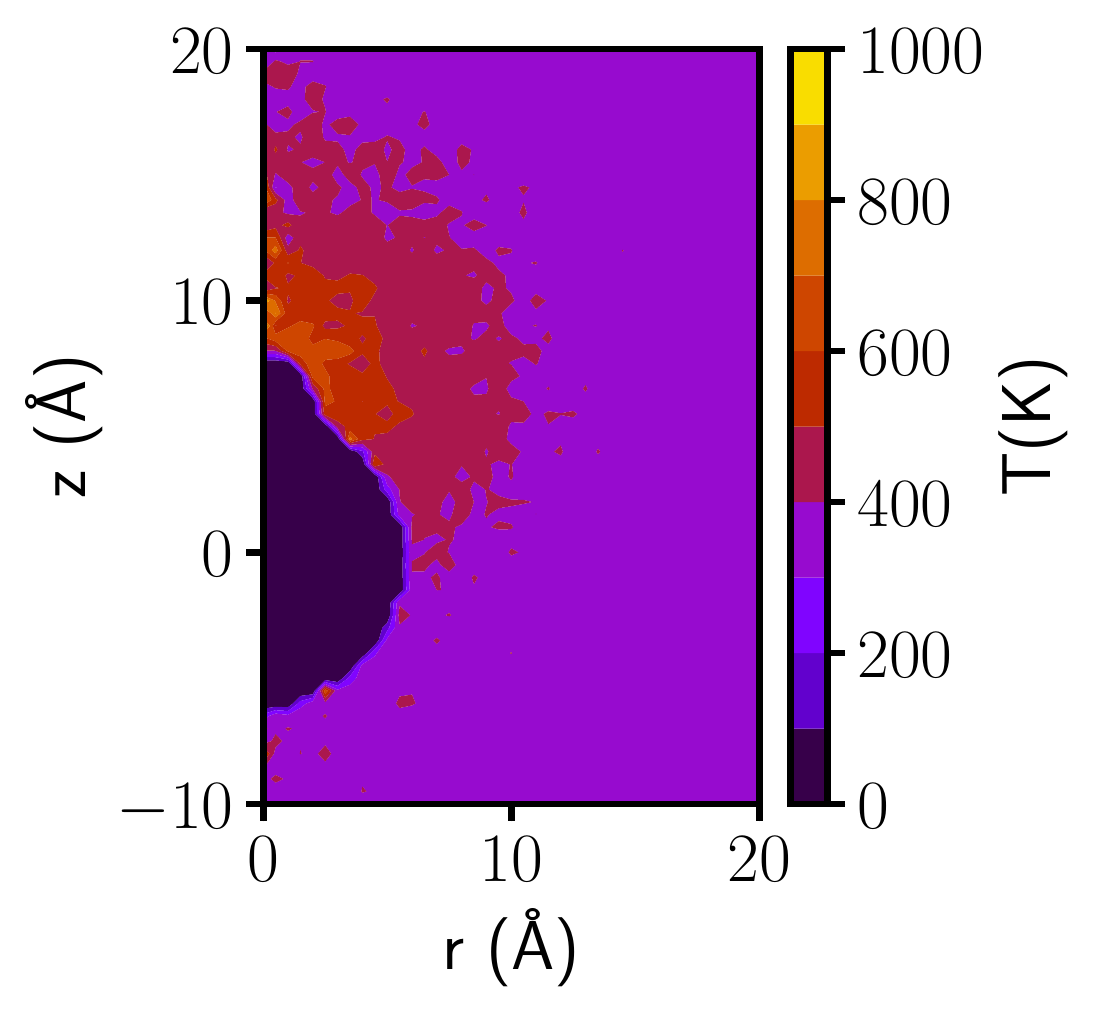}
\caption{Distribution of water temperature produced by the periodic inversion of the C$_{60}$-X dipole direction ($\tau = 0.5$ ps). Cylindrical coordinates are employed, with a radial distance $r$ to the axis defined by the center of the C$_{60}$ molecule and the X atom, and a height $z$ with origin at the center of the C$_{60}$ molecule. The dark region corresponds to the location of the C$_{60}$-X molecule, with the X atom located at $r=0$ and $z=5.96$ {\AA}.}
\label{fig:temperature}
\end{figure}

Further insight can be gained from the temperature distribution around C$_{60}$-X, Fig.~\ref{fig:temperature}. Periodic switching of the C-X dipole results in substantial temperature gradients, albeit limited to its immediate environment: the strong dampening effect of water's high thermal conductivity results in almost constant temperature at distances beyond 20 \AA. (Despite such steep decay, a global temperature drift might still be an issue. It has been a  goal of our simulations to ensure that enhanced diffusion is not due to overall temperature increase~\cite{golestanian:2015}. This is guaranteed at the expense of using large systems, with $\approx 260,000$ water molecules, i.e. a 200 {\AA} box side). 
The inhomogeneous distribution of temperature around C$_{60}$-X indicates that heat transfer generated at the solute is spread anisotropically.
As noted in the analysis of enzyme motion~\cite{golestanian:2015}, it is not obvious that excess heat should be channeled in a definite direction, isotropic spreading should be the norm instead. In the present case, there is a strong coupling between the pulsating dipole and the immediate water molecules, and thus energy is overwhelmingly transferred into their direction~\cite{rey:2015}. The rest of water molecules solvating the fullerene are screened and respond feebly. To summarize, directionality (and thus inhomogeneous temperature distribution) is achieved here by the inhomogeneous distribution of accepting modes (water librations). 

The propulsion speeds obtained, shown in Fig.~\ref{fig:speed}, are extremely high ($\sim 10$ m/s) as compared to usual values for microswimmers ($\sim 10 \; \mu$m/s). The feasibility of attaining such speeds rests upon the ability to transfer the required power to the fluorophore. To estimate whether such powers can be achieved in a realistic system we focus on coumarin 153 (Cu153), a small dye which could play the role of the C-X dipole. Cu153 has been extensively studied in solvation relaxation studies~\cite{maroncelli:1987,lewis:1998}, it is characterized by a transition dipole moment of $\mu \sim 5$ D and a fluorescence time of $\tau_{rlx} \sim 5 $ ns, both typical of fluorescent molecules~\cite{lakowicz:2006}. In an experimental setting, two lasers would be required, one with a frequency centered at the maximum of Cu153 absorption band  $\nu_{abs}$ (driving absorption) and the other one centered at its emission band  $\nu_{em}$ (driving stimulated emission). 

We can estimate the power required to move the population between the ground state and the excited state of Cu153 in a time $\tau$. Given the strong coupling to the solvent, transitions occur over a range of frequencies and the transition rate $\Gamma$ can be calculated as ~\cite{Strickler:1962} 
\begin{equation}
\Gamma = \frac{2303c}{hn N_A}\int d\nu \   \rho(\nu) \epsilon(\nu)/\nu ,     
\label{eq:nn}
\end{equation}
where $\rho$ is the radiation density, $n$ the refractive index of the medium, $c$ the velocity of light in vacuum, $N_A$ Avogadro's number, and $\epsilon$ the molar extinction coefficient (in units of M$^{-1}$cm$^{-1}$). Assuming the laser emission spectrum is narrow compared to the breadth of the absorption profile (about 4000 cm$^{-1}$ for Cu153 in a polar solvent~\cite{Li:2016}), $\epsilon(\nu)$ can be replaced by its value at the maximum of the absorption band $\nu_{abs} =$ 24000 cm$^{-1}$, $\epsilon(\nu_{abs}) \approx$ 21000 M$^{-1}$cm$^{-1}$~\cite{dobek:2011}.  
Taking $n = 1$, the constant terms can be combined to give $ A = \frac{2303}{h\nu_{abs} N_A}   \epsilon(\nu_e) \approx 1.7 \times 10^5 \  {\rm cm}^2/{\rm J}$.  Eq.~(\ref{eq:nn}) now reads $\Gamma = AP$, with $P\equiv c\int d\nu \ \rho(\nu) $ being the integrated laser power in W/cm$^2$. Assuming a beam cross section of 1 cm$^2$ one finds that power required to excite half the ground state population every , e. g. $\tau$ = 10 ps, is given by $P=\ln(2)/A\tau \sim 4.1\times 10^5 $ W. A second laser with similar power output, with a frequency centered at the maximum of the emission band $\nu_{em} \approx $ 20000 cm$^{-1}$ is required for stimulated emission~\cite{dobek:2011}. 
Such powers can be obtained with 4.1 W lasers with a 1 kHz repetition rate and a 10 ns pulse duration.  
To estimate the speed that could be attained, using Cu153 under such laser radiation, we resort to the linear dependence found between speed and power absorbed, see Fig.~\ref{fig:speed}. Considering that the transition dipole for Cu153 is $\approx 5$ times smaller than that for C-X, and that only a quarter of the population is excited at a time, the power absorbed by Cu153 should be about 20 times smaller than for the C-X dipole, inducing a still substantial speed of $\approx 0.25$~m/s.

A crucial issue, connected with the input power required, is whether propulsion might be thwarted by optical trapping, an effect not explicitly taken into account in our simulations. In fact, the first experimental demonstration of self-thermophoresis suffered from this limitation~\cite{jiang:2010}. 
The impossibility to avoid optical trapping for micrometer particles probably explains why self-thermophoresis has received limited attention~\cite{bechinger:2016}. The case considered here belongs to the Rayleigh regime (radiation wavelength much larger than the particle dimension), which has been extensively studied at near-resonant conditions~\cite{agayan:2002}. No trapping is possible for radiation in exact resonance, in stark contrast with half-metallic particles \cite{jiang:2010}. In addition, although optical trapping is enhanced for slightly off-resonance radiation, its effect is inversely proportional to the particle's volume~\cite{agayan:2002} and huge powers are required to optically confine nanometric particles. Indeed, intensities as high as $10^{9}$ W/cm$^2$ are required for dielectric particles of 14 nm size (still one order of magnitude larger than our swimmer) in {\em water}~\cite{ashkin:1986}, much larger than the intensities needed to induce significant propulsion of our particles.

\begin{figure}[ht]
\includegraphics[height=4cm]{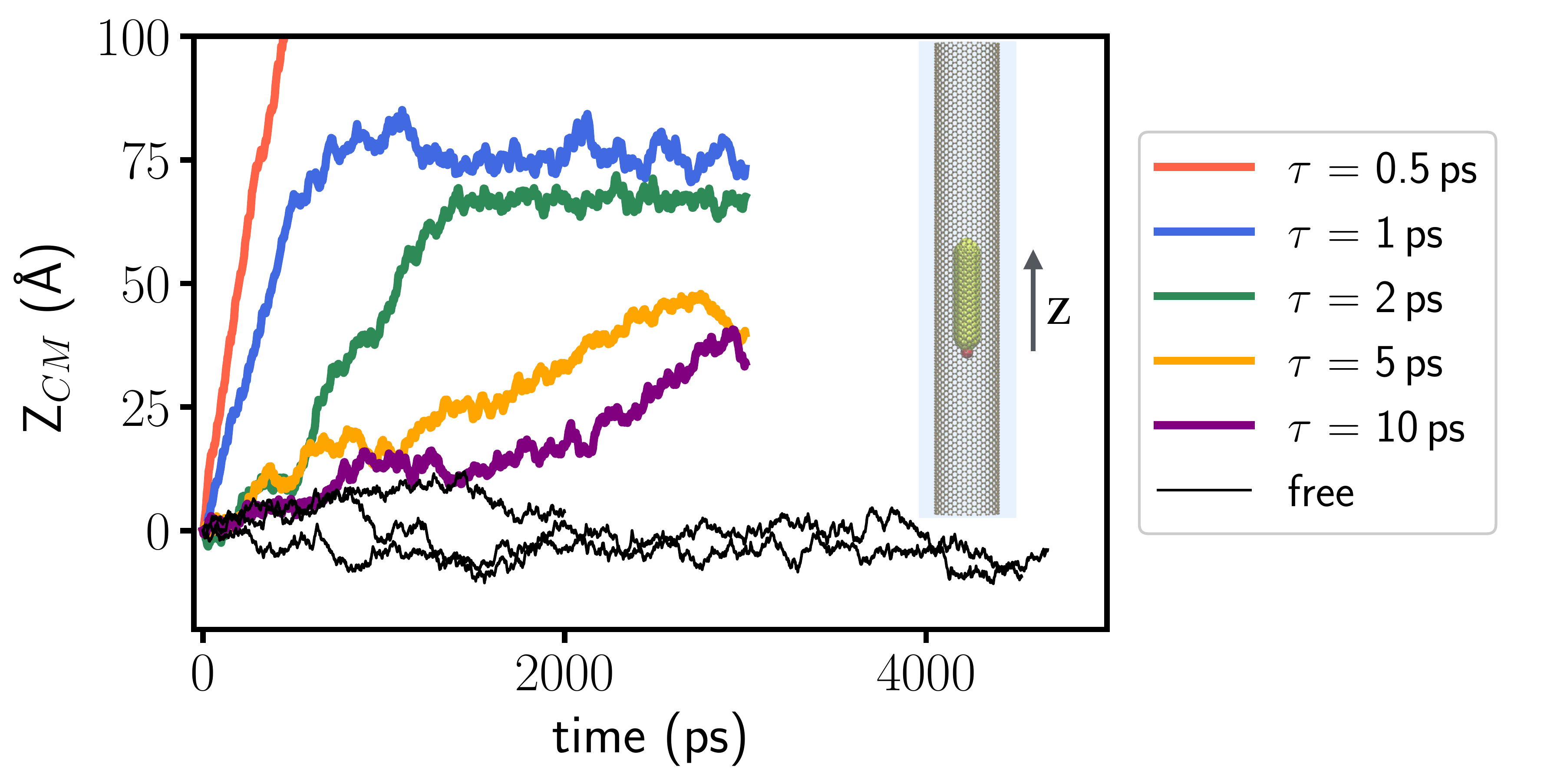}
\caption{Net displacement in time of the center of mass of the CNT-X molecule along the axis direction of the confining nanotube. Thin black lines correspond to non-actuated cases in which the dynamics is diffusive, whereas thick colored lines represent the results obtained for cases actuated by periodically inverting the CNT-X dipole direction with different switching periods $\tau$.}
\label{fig:nanotube}
\end{figure}

We now turn to a pervasive problem for microswimmers. The dynamics of the actuated C$_{60}$-X molecule is intrinsically Brownian, and directed motion only persists for limited intervals, characterized by the persistence length $\lambda \sim v\cdot \tau_r$, with $ \tau_r \equiv D_r^{-1}$~\cite{bechinger:2016}. Since $\tau_r \sim R^3$, as the radius is reduced the persistence length diminishes, motion becomes more random and less directed. While steering the propulsion in the direction of interest is thus a tall order in the bulk, we now demonstrate that it can be achieved in a restricted environment. In particular, we test the propulsion mechanism in the interior of a long and narrow nanotube submerged in water using all-atom MD simulations~\cite{SM}. The idea is to force its propulsion along the nanotube by restricting its rotation, which is accomplished using an elongated swimmer (as opposed to the spherical shape used so far). We use a relative of C$_{60}$-X, namely a 4nm long capped (5,5)-carbon nanotube with an atom X attached at one end, a nanoswimmer that we call CNT-X~\cite{SM} (see inset of Fig.~\ref{fig:nanotube}).

Repeating the same procedure of actuation (periodic dipole inversion) as for C$_{60}$-X, Fig.~\ref{fig:nanotube} shows the progress of the center of mass of CNT-X as it moves along the confining nanotube (the X moiety being located at its lower end). The trajectories displayed show the extraordinary degree of control attained: once the pumping starts, and irrespective of the power applied, the CNT-X particle moves along the nanotube chasing its own lower temperature end. Further analysis of the simulations for C$_{60}$-X in bulk water confirms that motion also takes place in the direction defined by the lower temperature hemisphere. Both cases (C$_{60}$-X, CNT-X) are therefore in line with the usual thermophobic behavior \cite{duhr:2006,piazza:2008}. 
The analysis of the CNT-X trajectories allows for a clear-cut estimation of its average speed (net displacement over the time elapsed), circumventing the ambiguities of multiparameter fits. The results are displayed in Fig.~\ref{fig:speed} side by side with those for C$_{60}$-X, showing similar magnitude and the same --linear-- dependence on input power. The speed similarity of C$_{60}$-X and CNT-X, despite the larger size of the latter, is probably related to the strong decrease of the effective viscosity of the solvent in confined environments~\cite{thomas:2008}. 

An additional unexpected feature is apparent in Fig.~\ref{fig:nanotube}: for low power, once CNT-X reaches the border of the confining nanotube it can get trapped at the nanotube-bulk boundary for times of at least several ns (horizontal sections of the blue and green curves), suggesting the idea that it might be possible to produce a pump-like mechanism. To this end, we have attached an additional atom (Y) at the opposite end (resulting in a Y-CNT-X molecule). We have now two propellers: the Y-C and C-X dipoles located at opposite ends of the CNT. They can be separately activated as each one might be characterized by a distinct absorption frequency, so that no focused radiation would be required. We have run trajectories where we have alternatively activated each propeller and succeeded in steering the Y-CNT-X nanoswimmer up and down the nanotube in a fully predictable way, without it ever escaping into the bulk (Video S1 in Supplemental Material~\cite{SM}. 

In conclusion, we have demonstrated the possibility to use light-induced solvation relaxation to propel nanometric particles in water. The solvent dynamics resulting from charge redistribution within a fluorophore (grafted to a nanoparticle) generates large temperature gradients, which induce propulsion towards the lower temperature side of the nanoparticle, with speed proportional to input power, consistent with self-thermophoresis. The proposed mechanism also works in confined environments and allows to direct the propulsion in a given direction. It is to be emphasized that, in contrast to other theoretical approaches describing phoretic effects, for the first time an all-atom model with off-the shelf interaction potentials has been employed, where gradients are not imposed explicitly to generate locomotion but emerge naturally from the solvent relaxation process. Our findings may be relevant for the transport of cargo at the nanoscale in aqueous media (e.g. drug delivery) and for the development of nanofluidic devices. It should be considered in addition that there is an ample variety of molecular excitation processes yet to be explored, and by which much larger energies could be deposited in the immediate environment of the nanoswimmer.

\begin{acknowledgements}

This work was supported by FIS2015-66879-C2-1-P (MINECO/FEDER)(RR). This work was partially funded by Horizon 2020 program through 766972-FET-OPEN-NANOPHLOW. ELS gratefully acknowledges support from NSF via Grant No. CHE-1566108.

\end{acknowledgements}


\begin{thebibliography}{46}%
\makeatletter
\providecommand \@ifxundefined [1]{%
 \@ifx{#1\undefined}
}%
\providecommand \@ifnum [1]{%
 \ifnum #1\expandafter \@firstoftwo
 \else \expandafter \@secondoftwo
 \fi
}%
\providecommand \@ifx [1]{%
 \ifx #1\expandafter \@firstoftwo
 \else \expandafter \@secondoftwo
 \fi
}%
\providecommand \natexlab [1]{#1}%
\providecommand \enquote  [1]{``#1''}%
\providecommand \bibnamefont  [1]{#1}%
\providecommand \bibfnamefont [1]{#1}%
\providecommand \citenamefont [1]{#1}%
\providecommand \href@noop [0]{\@secondoftwo}%
\providecommand \href [0]{\begingroup \@sanitize@url \@href}%
\providecommand \@href[1]{\@@startlink{#1}\@@href}%
\providecommand \@@href[1]{\endgroup#1\@@endlink}%
\providecommand \@sanitize@url [0]{\catcode `\\12\catcode `\$12\catcode
  `\&12\catcode `\#12\catcode `\^12\catcode `\_12\catcode `\%12\relax}%
\providecommand \@@startlink[1]{}%
\providecommand \@@endlink[0]{}%
\providecommand \url  [0]{\begingroup\@sanitize@url \@url }%
\providecommand \@url [1]{\endgroup\@href {#1}{\urlprefix }}%
\providecommand \urlprefix  [0]{URL }%
\providecommand \Eprint [0]{\href }%
\providecommand \doibase [0]{http://dx.doi.org/}%
\providecommand \selectlanguage [0]{\@gobble}%
\providecommand \bibinfo  [0]{\@secondoftwo}%
\providecommand \bibfield  [0]{\@secondoftwo}%
\providecommand \translation [1]{[#1]}%
\providecommand \BibitemOpen [0]{}%
\providecommand \bibitemStop [0]{}%
\providecommand \bibitemNoStop [0]{.\EOS\space}%
\providecommand \EOS [0]{\spacefactor3000\relax}%
\providecommand \BibitemShut  [1]{\csname bibitem#1\endcsname}%
\let\auto@bib@innerbib\@empty
\bibitem [{\citenamefont {Purcell}(1977)}]{purcell:1977}%
  \BibitemOpen
  \bibfield  {author} {\bibinfo {author} {\bibfnamefont {E.~M.}\ \bibnamefont
  {Purcell}},\ }\href@noop {} {\bibfield  {journal} {\bibinfo  {journal} {Am.
  J. Phys.}\ }\textbf {\bibinfo {volume} {45}},\ \bibinfo {pages} {3} (\bibinfo
  {year} {1977})}\BibitemShut {NoStop}%
\bibitem [{\citenamefont {Kapral}(2013)}]{kapral:2013}%
  \BibitemOpen
  \bibfield  {author} {\bibinfo {author} {\bibfnamefont {R.}~\bibnamefont
  {Kapral}},\ }\href@noop {} {\bibfield  {journal} {\bibinfo  {journal} {J.
  Chem. Phys.}\ }\textbf {\bibinfo {volume} {138}},\ \bibinfo {pages} {020901}
  (\bibinfo {year} {2013})}\BibitemShut {NoStop}%
\bibitem [{\citenamefont {Elgeti}\ \emph {et~al.}(2015)\citenamefont {Elgeti},
  \citenamefont {Winkler},\ and\ \citenamefont {Gompper}}]{elgeti:2015}%
  \BibitemOpen
  \bibfield  {author} {\bibinfo {author} {\bibfnamefont {J.}~\bibnamefont
  {Elgeti}}, \bibinfo {author} {\bibfnamefont {R.~G.}\ \bibnamefont {Winkler}},
  \ and\ \bibinfo {author} {\bibfnamefont {G.}~\bibnamefont {Gompper}},\
  }\href@noop {} {\bibfield  {journal} {\bibinfo  {journal} {Rep. Prog. Phys.}\
  }\textbf {\bibinfo {volume} {78}},\ \bibinfo {pages} {056601} (\bibinfo
  {year} {2015})}\BibitemShut {NoStop}%
\bibitem [{\citenamefont {Bechinger}\ \emph {et~al.}(2016)\citenamefont
  {Bechinger}, \citenamefont {Di~Leonardo}, \citenamefont {L{\"o}wen},
  \citenamefont {Reichhardt}, \citenamefont {Volpe},\ and\ \citenamefont
  {Volpe}}]{bechinger:2016}%
  \BibitemOpen
  \bibfield  {author} {\bibinfo {author} {\bibfnamefont {C.}~\bibnamefont
  {Bechinger}}, \bibinfo {author} {\bibfnamefont {R.}~\bibnamefont
  {Di~Leonardo}}, \bibinfo {author} {\bibfnamefont {H.}~\bibnamefont
  {L{\"o}wen}}, \bibinfo {author} {\bibfnamefont {C.}~\bibnamefont
  {Reichhardt}}, \bibinfo {author} {\bibfnamefont {G.}~\bibnamefont {Volpe}}, \
  and\ \bibinfo {author} {\bibfnamefont {G.}~\bibnamefont {Volpe}},\
  }\href@noop {} {\bibfield  {journal} {\bibinfo  {journal} {Rev. Mod. Phys.}\
  }\textbf {\bibinfo {volume} {88}},\ \bibinfo {pages} {045006} (\bibinfo
  {year} {2016})}\BibitemShut {NoStop}%
\bibitem [{\citenamefont {Moran}\ and\ \citenamefont
  {Posner}(2017)}]{moran:2017}%
  \BibitemOpen
  \bibfield  {author} {\bibinfo {author} {\bibfnamefont {J.~L.}\ \bibnamefont
  {Moran}}\ and\ \bibinfo {author} {\bibfnamefont {J.~D.}\ \bibnamefont
  {Posner}},\ }\href@noop {} {\bibfield  {journal} {\bibinfo  {journal} {Annu.
  Rev. Fluid Mech.}\ }\textbf {\bibinfo {volume} {49}},\ \bibinfo {pages} {511}
  (\bibinfo {year} {2017})}\BibitemShut {NoStop}%
\bibitem [{\citenamefont {Xu}\ \emph {et~al.}(2017{\natexlab{a}})\citenamefont
  {Xu}, \citenamefont {Mou}, \citenamefont {Gong}, \citenamefont {Luo},\ and\
  \citenamefont {Guan}}]{xu:2017}%
  \BibitemOpen
  \bibfield  {author} {\bibinfo {author} {\bibfnamefont {L.}~\bibnamefont
  {Xu}}, \bibinfo {author} {\bibfnamefont {F.}~\bibnamefont {Mou}}, \bibinfo
  {author} {\bibfnamefont {H.}~\bibnamefont {Gong}}, \bibinfo {author}
  {\bibfnamefont {M.}~\bibnamefont {Luo}}, \ and\ \bibinfo {author}
  {\bibfnamefont {J.}~\bibnamefont {Guan}},\ }\href@noop {} {\bibfield
  {journal} {\bibinfo  {journal} {Chem. Soc. Rev.}\ }\textbf {\bibinfo {volume}
  {46}},\ \bibinfo {pages} {6905} (\bibinfo {year}
  {2017}{\natexlab{a}})}\BibitemShut {NoStop}%
\bibitem [{\citenamefont {Xu}\ \emph {et~al.}(2017{\natexlab{b}})\citenamefont
  {Xu}, \citenamefont {Gao}, \citenamefont {Xu}, \citenamefont {Zhang},\ and\
  \citenamefont {Wang}}]{xu:2017bis}%
  \BibitemOpen
  \bibfield  {author} {\bibinfo {author} {\bibfnamefont {T.}~\bibnamefont
  {Xu}}, \bibinfo {author} {\bibfnamefont {W.}~\bibnamefont {Gao}}, \bibinfo
  {author} {\bibfnamefont {L.-P.}\ \bibnamefont {Xu}}, \bibinfo {author}
  {\bibfnamefont {X.}~\bibnamefont {Zhang}}, \ and\ \bibinfo {author}
  {\bibfnamefont {S.}~\bibnamefont {Wang}},\ }\href@noop {} {\bibfield
  {journal} {\bibinfo  {journal} {Adv. Mater.}\ }\textbf {\bibinfo {volume}
  {29}},\ \bibinfo {pages} {1603250} (\bibinfo {year}
  {2017}{\natexlab{b}})}\BibitemShut {NoStop}%
\bibitem [{\citenamefont {Garc{\'\i}a-Torres}\ \emph
  {et~al.}(2018)\citenamefont {Garc{\'\i}a-Torres}, \citenamefont {Calero},
  \citenamefont {Sagu{\'e}s}, \citenamefont {Pagonabarraga},\ and\
  \citenamefont {Tierno}}]{calero:2018}%
  \BibitemOpen
  \bibfield  {author} {\bibinfo {author} {\bibfnamefont {J.}~\bibnamefont
  {Garc{\'\i}a-Torres}}, \bibinfo {author} {\bibfnamefont {C.}~\bibnamefont
  {Calero}}, \bibinfo {author} {\bibfnamefont {F.}~\bibnamefont {Sagu{\'e}s}},
  \bibinfo {author} {\bibfnamefont {I.}~\bibnamefont {Pagonabarraga}}, \ and\
  \bibinfo {author} {\bibfnamefont {P.}~\bibnamefont {Tierno}},\ }\href@noop {}
  {\bibfield  {journal} {\bibinfo  {journal} {Nat. Commun.}\ }\textbf {\bibinfo
  {volume} {9}},\ \bibinfo {pages} {1663} (\bibinfo {year} {2018})}\BibitemShut
  {NoStop}%
\bibitem [{\citenamefont {Jiang}\ \emph {et~al.}(2010)\citenamefont {Jiang},
  \citenamefont {Yoshinaga},\ and\ \citenamefont {Sano}}]{jiang:2010}%
  \BibitemOpen
  \bibfield  {author} {\bibinfo {author} {\bibfnamefont {H.-R.}\ \bibnamefont
  {Jiang}}, \bibinfo {author} {\bibfnamefont {N.}~\bibnamefont {Yoshinaga}}, \
  and\ \bibinfo {author} {\bibfnamefont {M.}~\bibnamefont {Sano}},\ }\href@noop
  {} {\bibfield  {journal} {\bibinfo  {journal} {Phys. Rev. Lett.}\ }\textbf
  {\bibinfo {volume} {105}},\ \bibinfo {pages} {268302} (\bibinfo {year}
  {2010})}\BibitemShut {NoStop}%
\bibitem [{\citenamefont {Yang}\ and\ \citenamefont
  {Ripoll}(2011)}]{yang:2011}%
  \BibitemOpen
  \bibfield  {author} {\bibinfo {author} {\bibfnamefont {M.}~\bibnamefont
  {Yang}}\ and\ \bibinfo {author} {\bibfnamefont {M.}~\bibnamefont {Ripoll}},\
  }\href@noop {} {\bibfield  {journal} {\bibinfo  {journal} {Phys. Rev. E}\
  }\textbf {\bibinfo {volume} {84}},\ \bibinfo {pages} {061401} (\bibinfo
  {year} {2011})}\BibitemShut {NoStop}%
\bibitem [{\citenamefont {Yardley}(2012)}]{yardley:1980}%
  \BibitemOpen
  \bibfield  {author} {\bibinfo {author} {\bibfnamefont {J.}~\bibnamefont
  {Yardley}},\ }\href@noop {} {\emph {\bibinfo {title} {Introduction to
  Molecular Energy Transfer}}}\ (\bibinfo  {publisher} {Elsevier},\ \bibinfo
  {year} {2012})\BibitemShut {NoStop}%
\bibitem [{\citenamefont {Oxtoby}(1981)}]{oxtoby:1981}%
  \BibitemOpen
  \bibfield  {author} {\bibinfo {author} {\bibfnamefont {D.~W.}\ \bibnamefont
  {Oxtoby}},\ }\href@noop {} {\bibfield  {journal} {\bibinfo  {journal} {Ann.
  Rev. Phys. Chem.}\ }\textbf {\bibinfo {volume} {32}},\ \bibinfo {pages} {77}
  (\bibinfo {year} {1981})}\BibitemShut {NoStop}%
\bibitem [{\citenamefont {Bagchi}\ \emph {et~al.}(1983)\citenamefont {Bagchi},
  \citenamefont {Fleming},\ and\ \citenamefont {Oxtoby}}]{bagchi:1983}%
  \BibitemOpen
  \bibfield  {author} {\bibinfo {author} {\bibfnamefont {B.}~\bibnamefont
  {Bagchi}}, \bibinfo {author} {\bibfnamefont {G.~R.}\ \bibnamefont {Fleming}},
  \ and\ \bibinfo {author} {\bibfnamefont {D.~W.}\ \bibnamefont {Oxtoby}},\
  }\href@noop {} {\bibfield  {journal} {\bibinfo  {journal} {J. Chem. Phys.}\
  }\textbf {\bibinfo {volume} {78}},\ \bibinfo {pages} {7375} (\bibinfo {year}
  {1983})}\BibitemShut {NoStop}%
\bibitem [{\citenamefont {Rey}\ and\ \citenamefont {Hynes}(1996)}]{rey:1996}%
  \BibitemOpen
  \bibfield  {author} {\bibinfo {author} {\bibfnamefont {R.}~\bibnamefont
  {Rey}}\ and\ \bibinfo {author} {\bibfnamefont {J.~T.}\ \bibnamefont
  {Hynes}},\ }\href@noop {} {\bibfield  {journal} {\bibinfo  {journal} {J.
  Chem. Phys.}\ }\textbf {\bibinfo {volume} {104}},\ \bibinfo {pages} {2356}
  (\bibinfo {year} {1996})}\BibitemShut {NoStop}%
\bibitem [{\citenamefont {Ohta}\ \emph {et~al.}(1999)\citenamefont {Ohta},
  \citenamefont {Kang}, \citenamefont {Tominaga},\ and\ \citenamefont
  {Yoshihara}}]{ohta1999}%
  \BibitemOpen
  \bibfield  {author} {\bibinfo {author} {\bibfnamefont {K.}~\bibnamefont
  {Ohta}}, \bibinfo {author} {\bibfnamefont {T.~J.}\ \bibnamefont {Kang}},
  \bibinfo {author} {\bibfnamefont {K.}~\bibnamefont {Tominaga}}, \ and\
  \bibinfo {author} {\bibfnamefont {K.}~\bibnamefont {Yoshihara}},\ }\href@noop
  {} {\bibfield  {journal} {\bibinfo  {journal} {Chem. Phys.}\ }\textbf
  {\bibinfo {volume} {242}},\ \bibinfo {pages} {103} (\bibinfo {year}
  {1999})}\BibitemShut {NoStop}%
\bibitem [{\citenamefont {Sibert~III}\ and\ \citenamefont
  {Rey}(2002)}]{sibert:2002}%
  \BibitemOpen
  \bibfield  {author} {\bibinfo {author} {\bibfnamefont {E.~L.}\ \bibnamefont
  {Sibert~III}}\ and\ \bibinfo {author} {\bibfnamefont {R.}~\bibnamefont
  {Rey}},\ }\href@noop {} {\bibfield  {journal} {\bibinfo  {journal} {J. Chem.
  Phys.}\ }\textbf {\bibinfo {volume} {116}},\ \bibinfo {pages} {237} (\bibinfo
  {year} {2002})}\BibitemShut {NoStop}%
\bibitem [{\citenamefont {Rey}\ \emph {et~al.}(2004)\citenamefont {Rey},
  \citenamefont {M{\o}ller},\ and\ \citenamefont {Hynes}}]{rey:2004}%
  \BibitemOpen
  \bibfield  {author} {\bibinfo {author} {\bibfnamefont {R.}~\bibnamefont
  {Rey}}, \bibinfo {author} {\bibfnamefont {K.~B.}\ \bibnamefont {M{\o}ller}},
  \ and\ \bibinfo {author} {\bibfnamefont {J.~T.}\ \bibnamefont {Hynes}},\
  }\href@noop {} {\bibfield  {journal} {\bibinfo  {journal} {Chem. Rev.}\
  }\textbf {\bibinfo {volume} {104}},\ \bibinfo {pages} {1915} (\bibinfo {year}
  {2004})}\BibitemShut {NoStop}%
\bibitem [{\citenamefont {Rey}\ and\ \citenamefont {Hynes}(2012)}]{rey:2012}%
  \BibitemOpen
  \bibfield  {author} {\bibinfo {author} {\bibfnamefont {R.}~\bibnamefont
  {Rey}}\ and\ \bibinfo {author} {\bibfnamefont {J.~T.}\ \bibnamefont
  {Hynes}},\ }\href@noop {} {\bibfield  {journal} {\bibinfo  {journal} {Phys.
  Chem. Chem. Phys.}\ }\textbf {\bibinfo {volume} {14}},\ \bibinfo {pages}
  {6332} (\bibinfo {year} {2012})}\BibitemShut {NoStop}%
\bibitem [{\citenamefont {Maroncelli}(1993)}]{maroncelli:1993}%
  \BibitemOpen
  \bibfield  {author} {\bibinfo {author} {\bibfnamefont {M.}~\bibnamefont
  {Maroncelli}},\ }\href@noop {} {\bibfield  {journal} {\bibinfo  {journal} {J.
  Mol. Liq.}\ }\textbf {\bibinfo {volume} {57}},\ \bibinfo {pages} {1}
  (\bibinfo {year} {1993})}\BibitemShut {NoStop}%
\bibitem [{\citenamefont {Bagchi}\ and\ \citenamefont
  {Jana}(2010)}]{bagchi:2010}%
  \BibitemOpen
  \bibfield  {author} {\bibinfo {author} {\bibfnamefont {B.}~\bibnamefont
  {Bagchi}}\ and\ \bibinfo {author} {\bibfnamefont {B.}~\bibnamefont {Jana}},\
  }\href@noop {} {\bibfield  {journal} {\bibinfo  {journal} {Chem. Soc. Rev.}\
  }\textbf {\bibinfo {volume} {39}},\ \bibinfo {pages} {1936} (\bibinfo {year}
  {2010})}\BibitemShut {NoStop}%
\bibitem [{\citenamefont {Rey}\ and\ \citenamefont {Hynes}(2015)}]{rey:2015}%
  \BibitemOpen
  \bibfield  {author} {\bibinfo {author} {\bibfnamefont {R.}~\bibnamefont
  {Rey}}\ and\ \bibinfo {author} {\bibfnamefont {J.~T.}\ \bibnamefont
  {Hynes}},\ }\href@noop {} {\bibfield  {journal} {\bibinfo  {journal} {J.
  Phys. Chem. B}\ }\textbf {\bibinfo {volume} {119}},\ \bibinfo {pages} {7558}
  (\bibinfo {year} {2015})}\BibitemShut {NoStop}%
\bibitem [{\citenamefont {Riedel}\ \emph {et~al.}(2015)\citenamefont {Riedel},
  \citenamefont {Gabizon}, \citenamefont {Wilson}, \citenamefont {Hamadani},
  \citenamefont {Tsekouras}, \citenamefont {Marqusee}, \citenamefont
  {Press{\'e}},\ and\ \citenamefont {Bustamante}}]{riedel:2014}%
  \BibitemOpen
  \bibfield  {author} {\bibinfo {author} {\bibfnamefont {C.}~\bibnamefont
  {Riedel}}, \bibinfo {author} {\bibfnamefont {R.}~\bibnamefont {Gabizon}},
  \bibinfo {author} {\bibfnamefont {C.~A.}\ \bibnamefont {Wilson}}, \bibinfo
  {author} {\bibfnamefont {K.}~\bibnamefont {Hamadani}}, \bibinfo {author}
  {\bibfnamefont {K.}~\bibnamefont {Tsekouras}}, \bibinfo {author}
  {\bibfnamefont {S.}~\bibnamefont {Marqusee}}, \bibinfo {author}
  {\bibfnamefont {S.}~\bibnamefont {Press{\'e}}}, \ and\ \bibinfo {author}
  {\bibfnamefont {C.}~\bibnamefont {Bustamante}},\ }\href@noop {} {\bibfield
  {journal} {\bibinfo  {journal} {Nature}\ }\textbf {\bibinfo {volume} {517}},\
  \bibinfo {pages} {227} (\bibinfo {year} {2015})}\BibitemShut {NoStop}%
\bibitem [{\citenamefont {Wand}(2015)}]{wand:2014}%
  \BibitemOpen
  \bibfield  {author} {\bibinfo {author} {\bibfnamefont {A.~J.}\ \bibnamefont
  {Wand}},\ }\href@noop {} {\bibfield  {journal} {\bibinfo  {journal} {Nature}\
  }\textbf {\bibinfo {volume} {517}},\ \bibinfo {pages} {149} (\bibinfo {year}
  {2015})}\BibitemShut {NoStop}%
\bibitem [{\citenamefont {Golestanian}(2015)}]{golestanian:2015}%
  \BibitemOpen
  \bibfield  {author} {\bibinfo {author} {\bibfnamefont {R.}~\bibnamefont
  {Golestanian}},\ }\href@noop {} {\bibfield  {journal} {\bibinfo  {journal}
  {Phys. Rev. Lett.}\ }\textbf {\bibinfo {volume} {115}},\ \bibinfo {pages}
  {108102} (\bibinfo {year} {2015})}\BibitemShut {NoStop}%
\bibitem [{\citenamefont {Goodarzi}\ \emph {et~al.}(2017)\citenamefont
  {Goodarzi}, \citenamefont {Da~Ros}, \citenamefont {Conde}, \citenamefont
  {Sefat},\ and\ \citenamefont {Mozafari}}]{goodarzil:2017}%
  \BibitemOpen
  \bibfield  {author} {\bibinfo {author} {\bibfnamefont {S.}~\bibnamefont
  {Goodarzi}}, \bibinfo {author} {\bibfnamefont {T.}~\bibnamefont {Da~Ros}},
  \bibinfo {author} {\bibfnamefont {J.}~\bibnamefont {Conde}}, \bibinfo
  {author} {\bibfnamefont {F.}~\bibnamefont {Sefat}}, \ and\ \bibinfo {author}
  {\bibfnamefont {M.}~\bibnamefont {Mozafari}},\ }\href@noop {} {\bibfield
  {journal} {\bibinfo  {journal} {Mater. Today}\ }\textbf {\bibinfo {volume}
  {20}},\ \bibinfo {pages} {460} (\bibinfo {year} {2017})}\BibitemShut
  {NoStop}%
\bibitem [{\citenamefont {Gonzalez}\ \emph {et~al.}(2017)\citenamefont
  {Gonzalez}, \citenamefont {Lujan},\ and\ \citenamefont
  {Beran}}]{gonzalez:2017}%
  \BibitemOpen
  \bibfield  {author} {\bibinfo {author} {\bibfnamefont {M.}~\bibnamefont
  {Gonzalez}}, \bibinfo {author} {\bibfnamefont {S.}~\bibnamefont {Lujan}}, \
  and\ \bibinfo {author} {\bibfnamefont {K.~A.}\ \bibnamefont {Beran}},\
  }\href@noop {} {\bibfield  {journal} {\bibinfo  {journal} {Comput. Theor.
  Chem.}\ }\textbf {\bibinfo {volume} {1119}},\ \bibinfo {pages} {32} (\bibinfo
  {year} {2017})}\BibitemShut {NoStop}%
\bibitem [{SM()}]{SM}%
  \BibitemOpen
  \href@noop {} {}\bibinfo {note} {See Supplemental Material at [] for videos,
  details of the simulations and analysis, and additional figures.}\BibitemShut
  {Stop}%
\bibitem [{\citenamefont {Bruehl}\ and\ \citenamefont
  {Hynes}(1992)}]{bruehl:1992}%
  \BibitemOpen
  \bibfield  {author} {\bibinfo {author} {\bibfnamefont {M.}~\bibnamefont
  {Bruehl}}\ and\ \bibinfo {author} {\bibfnamefont {J.~T.}\ \bibnamefont
  {Hynes}},\ }\href@noop {} {\bibfield  {journal} {\bibinfo  {journal} {J.
  Phys. Chem.}\ }\textbf {\bibinfo {volume} {96}},\ \bibinfo {pages} {4068}
  (\bibinfo {year} {1992})}\BibitemShut {NoStop}%
\bibitem [{\citenamefont {Schile}\ and\ \citenamefont
  {Thompson}(2017)}]{schile:2017}%
  \BibitemOpen
  \bibfield  {author} {\bibinfo {author} {\bibfnamefont {A.~J.}\ \bibnamefont
  {Schile}}\ and\ \bibinfo {author} {\bibfnamefont {W.~H.}\ \bibnamefont
  {Thompson}},\ }\href@noop {} {\bibfield  {journal} {\bibinfo  {journal} {J.
  Chem. Phys.}\ }\textbf {\bibinfo {volume} {146}},\ \bibinfo {pages} {154109}
  (\bibinfo {year} {2017})}\BibitemShut {NoStop}%
\bibitem [{\citenamefont {Hynes}\ \emph {et~al.}(1979)\citenamefont {Hynes},
  \citenamefont {Kapral},\ and\ \citenamefont {Weinberg}}]{hynes:1979}%
  \BibitemOpen
  \bibfield  {author} {\bibinfo {author} {\bibfnamefont {J.~T.}\ \bibnamefont
  {Hynes}}, \bibinfo {author} {\bibfnamefont {R.}~\bibnamefont {Kapral}}, \
  and\ \bibinfo {author} {\bibfnamefont {M.}~\bibnamefont {Weinberg}},\
  }\href@noop {} {\bibfield  {journal} {\bibinfo  {journal} {J. Chem. Phys.}\
  }\textbf {\bibinfo {volume} {70}},\ \bibinfo {pages} {1456} (\bibinfo {year}
  {1979})}\BibitemShut {NoStop}%
\bibitem [{\citenamefont {Lai}\ \emph {et~al.}(2010)\citenamefont {Lai},
  \citenamefont {Kalweit},\ and\ \citenamefont {Drikakis}}]{lai:2010}%
  \BibitemOpen
  \bibfield  {author} {\bibinfo {author} {\bibfnamefont {M.}~\bibnamefont
  {Lai}}, \bibinfo {author} {\bibfnamefont {M.}~\bibnamefont {Kalweit}}, \ and\
  \bibinfo {author} {\bibfnamefont {D.}~\bibnamefont {Drikakis}},\ }\href@noop
  {} {\bibfield  {journal} {\bibinfo  {journal} {Mol. Sim.}\ }\textbf {\bibinfo
  {volume} {36}},\ \bibinfo {pages} {801} (\bibinfo {year} {2010})}\BibitemShut
  {NoStop}%
\bibitem [{\citenamefont {Rey}\ and\ \citenamefont {Hynes}(2016)}]{rey:2016}%
  \BibitemOpen
  \bibfield  {author} {\bibinfo {author} {\bibfnamefont {R.}~\bibnamefont
  {Rey}}\ and\ \bibinfo {author} {\bibfnamefont {J.~T.}\ \bibnamefont
  {Hynes}},\ }\href@noop {} {\bibfield  {journal} {\bibinfo  {journal} {J.
  Phys. Chem. B}\ }\textbf {\bibinfo {volume} {120}},\ \bibinfo {pages} {11287}
  (\bibinfo {year} {2016})}\BibitemShut {NoStop}%
\bibitem [{\citenamefont {ten Hagen}\ \emph {et~al.}(2011)\citenamefont {ten
  Hagen}, \citenamefont {van Teeffelen},\ and\ \citenamefont
  {L{\"o}wen}}]{hagen:2011}%
  \BibitemOpen
  \bibfield  {author} {\bibinfo {author} {\bibfnamefont {B.}~\bibnamefont {ten
  Hagen}}, \bibinfo {author} {\bibfnamefont {S.}~\bibnamefont {van Teeffelen}},
  \ and\ \bibinfo {author} {\bibfnamefont {H.}~\bibnamefont {L{\"o}wen}},\
  }\href@noop {} {\bibfield  {journal} {\bibinfo  {journal} {J. Phys. Condens.
  Matter}\ }\textbf {\bibinfo {volume} {23}},\ \bibinfo {pages} {194119}
  (\bibinfo {year} {2011})}\BibitemShut {NoStop}%
\bibitem [{\citenamefont {Golestanian}\ \emph {et~al.}(2007)\citenamefont
  {Golestanian}, \citenamefont {Liverpool},\ and\ \citenamefont
  {Ajdari}}]{golestanian:2007}%
  \BibitemOpen
  \bibfield  {author} {\bibinfo {author} {\bibfnamefont {R.}~\bibnamefont
  {Golestanian}}, \bibinfo {author} {\bibfnamefont {T.}~\bibnamefont
  {Liverpool}}, \ and\ \bibinfo {author} {\bibfnamefont {A.}~\bibnamefont
  {Ajdari}},\ }\href@noop {} {\bibfield  {journal} {\bibinfo  {journal} {New J.
  Phys.}\ }\textbf {\bibinfo {volume} {9}},\ \bibinfo {pages} {126} (\bibinfo
  {year} {2007})}\BibitemShut {NoStop}%
\bibitem [{\citenamefont {Sirk}\ \emph {et~al.}(2013)\citenamefont {Sirk},
  \citenamefont {Moore},\ and\ \citenamefont {Brown}}]{sirk:2013}%
  \BibitemOpen
  \bibfield  {author} {\bibinfo {author} {\bibfnamefont {T.~W.}\ \bibnamefont
  {Sirk}}, \bibinfo {author} {\bibfnamefont {S.}~\bibnamefont {Moore}}, \ and\
  \bibinfo {author} {\bibfnamefont {E.~F.}\ \bibnamefont {Brown}},\ }\href@noop
  {} {\bibfield  {journal} {\bibinfo  {journal} {J. Chem. Phys.}\ }\textbf
  {\bibinfo {volume} {138}},\ \bibinfo {pages} {064505} (\bibinfo {year}
  {2013})}\BibitemShut {NoStop}%
\bibitem [{\citenamefont {Piazza}\ and\ \citenamefont
  {Parola}(2008)}]{piazza:2008}%
  \BibitemOpen
  \bibfield  {author} {\bibinfo {author} {\bibfnamefont {R.}~\bibnamefont
  {Piazza}}\ and\ \bibinfo {author} {\bibfnamefont {A.}~\bibnamefont
  {Parola}},\ }\href@noop {} {\bibfield  {journal} {\bibinfo  {journal} {J.
  Phys. Condens. Matter}\ }\textbf {\bibinfo {volume} {20}},\ \bibinfo {pages}
  {153102} (\bibinfo {year} {2008})}\BibitemShut {NoStop}%
\bibitem [{\citenamefont {Maroncelli}\ and\ \citenamefont
  {Fleming}(1987)}]{maroncelli:1987}%
  \BibitemOpen
  \bibfield  {author} {\bibinfo {author} {\bibfnamefont {M.}~\bibnamefont
  {Maroncelli}}\ and\ \bibinfo {author} {\bibfnamefont {G.~R.}\ \bibnamefont
  {Fleming}},\ }\href {\doibase 10.1063/1.452460} {\bibfield  {journal}
  {\bibinfo  {journal} {J. Chem. Phys.}\ }\textbf {\bibinfo {volume} {86}},\
  \bibinfo {pages} {6221} (\bibinfo {year} {1987})},\ \Eprint
  {http://arxiv.org/abs/https://doi.org/10.1063/1.452460}
  {https://doi.org/10.1063/1.452460} \BibitemShut {NoStop}%
\bibitem [{\citenamefont {Lewis}\ and\ \citenamefont
  {Maroncelli}(1998)}]{lewis:1998}%
  \BibitemOpen
  \bibfield  {author} {\bibinfo {author} {\bibfnamefont {J.}~\bibnamefont
  {Lewis}}\ and\ \bibinfo {author} {\bibfnamefont {M.}~\bibnamefont
  {Maroncelli}},\ }\href@noop {} {\bibfield  {journal} {\bibinfo  {journal}
  {Chem. Phys. Lett.}\ }\textbf {\bibinfo {volume} {282}},\ \bibinfo {pages}
  {197} (\bibinfo {year} {1998})}\BibitemShut {NoStop}%
\bibitem [{\citenamefont {Lakowicz}(2013)}]{lakowicz:2006}%
  \BibitemOpen
  \bibfield  {author} {\bibinfo {author} {\bibfnamefont {J.~R.}\ \bibnamefont
  {Lakowicz}},\ }\href@noop {} {\emph {\bibinfo {title} {Principles of
  Fluorescence Spectroscopy}}}\ (\bibinfo  {publisher} {Springer Science \&
  Business Media},\ \bibinfo {year} {2013})\BibitemShut {NoStop}%
\bibitem [{\citenamefont {Strickler}\ and\ \citenamefont
  {Berg}(1962)}]{Strickler:1962}%
  \BibitemOpen
  \bibfield  {author} {\bibinfo {author} {\bibfnamefont {S.}~\bibnamefont
  {Strickler}}\ and\ \bibinfo {author} {\bibfnamefont {R.~A.}\ \bibnamefont
  {Berg}},\ }\href@noop {} {\bibfield  {journal} {\bibinfo  {journal} {J. Chem.
  Phys.}\ }\textbf {\bibinfo {volume} {37}},\ \bibinfo {pages} {814} (\bibinfo
  {year} {1962})}\BibitemShut {NoStop}%
\bibitem [{\citenamefont {Li}\ \emph {et~al.}(2016)\citenamefont {Li},
  \citenamefont {Yu},\ and\ \citenamefont {Lu}}]{Li:2016}%
  \BibitemOpen
  \bibfield  {author} {\bibinfo {author} {\bibfnamefont {S.}~\bibnamefont
  {Li}}, \bibinfo {author} {\bibfnamefont {A.}~\bibnamefont {Yu}}, \ and\
  \bibinfo {author} {\bibfnamefont {R.}~\bibnamefont {Lu}},\ }\href@noop {}
  {\bibfield  {journal} {\bibinfo  {journal} {Spectrochim. Acta A}\ }\textbf
  {\bibinfo {volume} {165}},\ \bibinfo {pages} {161} (\bibinfo {year}
  {2016})}\BibitemShut {NoStop}%
\bibitem [{\citenamefont {Dobek}(2011)}]{dobek:2011}%
  \BibitemOpen
  \bibfield  {author} {\bibinfo {author} {\bibfnamefont {K.}~\bibnamefont
  {Dobek}},\ }\href@noop {} {\bibfield  {journal} {\bibinfo  {journal} {J.
  Fluoresc.}\ }\textbf {\bibinfo {volume} {21}},\ \bibinfo {pages} {1547}
  (\bibinfo {year} {2011})}\BibitemShut {NoStop}%
\bibitem [{\citenamefont {Agayan}\ \emph {et~al.}(2002)\citenamefont {Agayan},
  \citenamefont {Gittes}, \citenamefont {Kopelman},\ and\ \citenamefont
  {Schmidt}}]{agayan:2002}%
  \BibitemOpen
  \bibfield  {author} {\bibinfo {author} {\bibfnamefont {R.~R.}\ \bibnamefont
  {Agayan}}, \bibinfo {author} {\bibfnamefont {F.}~\bibnamefont {Gittes}},
  \bibinfo {author} {\bibfnamefont {R.}~\bibnamefont {Kopelman}}, \ and\
  \bibinfo {author} {\bibfnamefont {C.~F.}\ \bibnamefont {Schmidt}},\
  }\href@noop {} {\bibfield  {journal} {\bibinfo  {journal} {Appl. Opt.}\
  }\textbf {\bibinfo {volume} {41}},\ \bibinfo {pages} {2318} (\bibinfo {year}
  {2002})}\BibitemShut {NoStop}%
\bibitem [{\citenamefont {Ashkin}\ \emph {et~al.}(1986)\citenamefont {Ashkin},
  \citenamefont {Dziedzic}, \citenamefont {Bjorkholm},\ and\ \citenamefont
  {Chu}}]{ashkin:1986}%
  \BibitemOpen
  \bibfield  {author} {\bibinfo {author} {\bibfnamefont {A.}~\bibnamefont
  {Ashkin}}, \bibinfo {author} {\bibfnamefont {J.~M.}\ \bibnamefont
  {Dziedzic}}, \bibinfo {author} {\bibfnamefont {J.}~\bibnamefont {Bjorkholm}},
  \ and\ \bibinfo {author} {\bibfnamefont {S.}~\bibnamefont {Chu}},\
  }\href@noop {} {\bibfield  {journal} {\bibinfo  {journal} {Opt. Lett.}\
  }\textbf {\bibinfo {volume} {11}},\ \bibinfo {pages} {288} (\bibinfo {year}
  {1986})}\BibitemShut {NoStop}%
\bibitem [{\citenamefont {Duhr}\ and\ \citenamefont {Braun}(2006)}]{duhr:2006}%
  \BibitemOpen
  \bibfield  {author} {\bibinfo {author} {\bibfnamefont {S.}~\bibnamefont
  {Duhr}}\ and\ \bibinfo {author} {\bibfnamefont {D.}~\bibnamefont {Braun}},\
  }\href@noop {} {\bibfield  {journal} {\bibinfo  {journal} {Proc. Natl. Acad.
  Sci.}\ }\textbf {\bibinfo {volume} {103}},\ \bibinfo {pages} {19678}
  (\bibinfo {year} {2006})}\BibitemShut {NoStop}%
\bibitem [{\citenamefont {Thomas}\ and\ \citenamefont
  {McGaughey}(2008)}]{thomas:2008}%
  \BibitemOpen
  \bibfield  {author} {\bibinfo {author} {\bibfnamefont {J.~A.}\ \bibnamefont
  {Thomas}}\ and\ \bibinfo {author} {\bibfnamefont {A.~J.}\ \bibnamefont
  {McGaughey}},\ }\href@noop {} {\bibfield  {journal} {\bibinfo  {journal}
  {Nano Lett.}\ }\textbf {\bibinfo {volume} {8}},\ \bibinfo {pages} {2788}
  (\bibinfo {year} {2008})}\BibitemShut {NoStop}%
\end{thebibliography}

\end{document}